\documentclass[11pt]{article}
\usepackage{epsf,amsmath,amssymb,graphicx,scalefnt}
\usepackage{rotating,ifthen,color,hyperref}
\newcommand{\bbhnnlo}{\href{http://www.robert-harlander.de/software/bbh@nnlo}{%
    \tt bbh@nnlo}}
\newcommand{\fs}[1]{{\abbrev #1FS}}

\newcommand{\pdf}{{\abbrev PDF}}
\newcommand{\muF}{\mu_\text{F}}
\newcommand{\muR}{\mu_\text{R}}
\newcommand{\mhiggs}{m_\text{H}}
\newcommand{\mbottom}{m_\text{b}}

\newcommand{\abbrev}{\scalefont{.9}}

\newcommand{\eqn}[1]{Eq.\,(\ref{#1})}
\newcommand{\fig}[1]{Fig.\,\ref{#1}}

\newcommand{\lhc}{{\abbrev LHC}}
\newcommand{\qcd}{{\abbrev QCD}}

\newcommand{\susy}{{\abbrev SUSY}}

\newcommand{\lo}{{\abbrev LO}}
\newcommand{\nlo}{{\abbrev NLO}}
\newcommand{\nnlo}{{\abbrev NNLO}}

\title{\vspace*{-6em}
  \begin{flushright}
    {\sf\small June 2011 --- CERN-PH-TH/2011-134 --
      FR-PHENO-2011-009 --
      TTK-11-17 -- WUB/11-04}
  \end{flushright}\vspace*{2em}
Bottom-quark associated Higgs-boson production:\\
reconciling the four- and five-flavour scheme approach}
\author{R.\ Harlander$^{1,2}$, M.\ Kr\"amer$^3$, and M.\ Schumacher$^4$\\[2em]
$^1$\it Fachbereich C, Bergische Universit\"at Wuppertal, 42097
  Wuppertal, Germany\\[2mm]
$^2$\it PH-TH, CERN, CH-1211, Geneva 23, Switzerland\\[2mm]
$^3$\it Institute for Theoretical Particle Physics and Cosmology\\ 
 \it RWTH Aachen University, D-52056 Aachen, Germany\\[2mm]
$^4$\it Fakult\"at f\"ur Mathematik und Physik, Albert-Ludwigs Universit\"at Freiburg,\\
\it D-79104 Freiburg, Germany}
\date{}
\begin{document}
\maketitle
\begin{abstract}
The main arguments in the discussion of the proper treatment of the
total inclusive cross section for bottom-quark associated Higgs-boson
production are briefly reviewed. A simple and pragmatic formula for the
combination of the so-called four- and five-flavour schemes is
suggested, including the treatment of the respective theory error
estimates. The numerical effects of this matching formula are discussed.
\end{abstract}



\section{Two approaches}\label{sec::intro}

The inclusive total cross section for bottom-quark associated
Higgs-boson production, denoted\footnote{The notation $(b\bar b)H$ is
  meant to indicate that the $b\bar b$ pair is not required as part of
  the signature in this process, so that its final state momenta must be
  integrated over the full phase space.}  $pp/p\bar p \to
(b\bar{b})H+X$, can be calculated in two different schemes. As the mass
of the bottom-quark is large compared to the \qcd{} scale, $\mbottom \gg
\Lambda_\text{\qcd}$, bottom-quark production is a perturbative process and
can be calculated order by order. Thus, in a four-flavour scheme
(\fs{4}), where one does not consider $b$-quarks as partons in the
proton, the lowest-order \qcd{} production processes are gluon-gluon
fusion and quark-antiquark annihilation, $gg \to b\bar{b}H$ and
$q\bar{q}\to b\bar{b}H$, respectively. However, the inclusive cross
section for $gg \to (b\bar{b})H$ develops logarithms of the form
$\ln(\muF/\mbottom)$, which arise from the splitting of gluons into
nearly collinear $b\bar{b}$ pairs. The large factorization scale $\muF
\approx \mhiggs/4$ corresponds to the upper limit of the collinear
region up to which factorization is
valid~\cite{Rainwater:2002hm,Plehn:2002vy,Maltoni:2003pn}. For
Higgs-boson masses $\mhiggs \gg 4 \mbottom$, the logarithms become large
and spoil the convergence of the perturbative series. The
$\ln(\muF/\mbottom)$ terms can be summed to all orders in perturbation
theory by introducing bottom parton densities. This defines the
so-called five-flavour scheme (\fs{5}). The use of bottom distribution
functions is based on the approximation that the outgoing $b$-quarks are
at small transverse momentum. In this scheme, the \lo{} process for the
inclusive $(b\bar{b})H$ cross section is bottom fusion, $b\bar{b} \to
H$.

If all orders in perturbation theory were taken into account, the four-
and five-flavour schemes would be identical, but the way of ordering the
perturbative expansion is different. At any finite order, the two
schemes include different parts of the all-order result, and the cross
section predictions do thus not match exactly. While this leads to an
ambiguity in the way the cross section is calculated, it also offers an
opportunity to test the importance of various higher-order terms and the
reliability of the theoretical prediction. The \fs{4} calculation is
available at \nlo{}~\cite{Dittmaier:2003ej,Dawson:2003kb}, while the
\fs{5} cross section has been calculated at \nnlo{}
accuracy~\cite{Harlander:2003ai}.  Electroweak corrections to the \fs{5}
process have been found to be small~\cite{Dittmaier:2006cz} and will not
be considered in the numerical results presented in this note.

In the next section we briefly 
summarize some of the features of the two schemes.


\section{Discussion}\label{sec::discussion}

In this section, we collect the main arguments and counter-arguments
that have been discussed in the physics community for and against either
the \fs{4} or the \fs{5} approach. Let us stress that at this point, we
consider only the total inclusive cross section $pp \to
(b\bar{b})H+X$. Once distributions or even tagged $b$-quarks are taken
into account, the discussion below has to be re-evaluated.

\begin{description}
\item[\it Gluon to $b$-quark splitting --- kinematical approximations
  and collinear logarithms:] In $pp$ or $p\bar p$ collisions,
  bottom-quarks arise at \lo{} in $\alpha_s$ through the splitting of a
  gluon. In the \fs{4}, this splitting is described by fixed-order
  perturbation theory, and includes the full dependence on the
  transverse momentum $p_{T}$ of the $b$-quark and its mass. In the
  \fs{5} scheme, the splitting arises by solving the {\abbrev DGLAP}
  evolution equations with five massless quark
  flavours~\cite{Barnett:1987jw}.

Integrating over phase space, the \fs{5} at \lo{} neglects all
contributions beyond $1/p_{T}^2$ as $p_{T}\to 0$, while they are
consistently taken into account in the \fs{4}. However, at higher
orders, the \fs{5} does include sub-leading finite $p_{T}$ effects; at
\nnlo{}, all the information of the \lo{} \fs{4} calculation is
incorporated in the \fs{5} as well. Note, however, that the \fs{4}
result is available through \nlo{}, so that the finite $p_{T}$ effects
are available one order higher than in the \fs{5} calculation.
  
   Since the factorization
  theorem holds only for massless quarks, in the \fs{5} scheme one has
  to assume $\mbottom=0$ (the bottom Yukawa coupling can -- and must --
  be kept finite though). As the only other scales of the inclusive
  cross section are $\mhiggs$ and the hadronic center-of-mass energy
  $s\gg \mhiggs^2$, the \fs{4} and the \fs{5} formally differ at order
  $\mbottom^2/\mhiggs^2$, even if all orders of the perturbation theory
  could be taken into account. The numerical size of these effects has
  been investigated in Ref.\,\cite{Buttar:2006zd} and was found to be
  negligible.

 In the opposite phase space region of very small transverse momenta of
 the $b$-quarks, the \fs{5} implicitly re-sums logarithmic terms of
 order $\alpha_s^{n+2}/p_{T}^2\cdot\ln^{2n-k}p_{T}^2$ ($n\geq 0$),
 where $k=0$ in the \lo{}, $k\in\{0,1\}$ in the \nlo{}, and
 $k\in\{0,1,2\}$ in the \nnlo{} calculation. In the \fs{4}, these
 effects are not re-summed, but taken into account only up to finite $n$.

\item[\it Renormalization/factorization scale dependence:] Due to the
  different complexity of the two \lo{} processes ($2\to3$ for the
  \fs{4} and $2\to 1$ for the \fs{5}), radiative corrections are known
  through \nnlo{} in the \fs{5}~\cite{Harlander:2003ai}, while they are
  available only through \nlo{} in the
  \fs{4}~\cite{Dittmaier:2003ej,Dawson:2003kb}. This results in a
  stronger dependence in the \fs{4} on the unphysical renormalization
  and factorization scales, $\muR$ and $\muF$, than in the \fs{5}.
\item[\it Top-quark induced effects:] Currently, the \fs{5} calculation
  neglects effects proportional to the top Yukawa
  coupling. Partly, they are taken into account in the gluon fusion
  induced cross section, $gg\to H+X$. What is missing are interference
  effects proportional to both the top and the bottom Yukawa
  coupling. In most models with an enhanced bottom Yukawa coupling (like
  \susy{} at large $\tan\beta$), the top Yukawa coupling is
  simultaneously suppressed though, and therefore these interference
  terms can be neglected. For consistency, these terms have been removed
  also from the \fs{4} calculations in the numerical results below.
\item[\it Parton distribution functions:] Technically, the transition
  from $n_f$ to $n_f+1$ flavours is well understood through
  \nnlo{}~\cite{Chuvakin:1999nx}. The exact value $\mu_\text{thr}$ where
  this transition is imposed is arbitrary (but fixed in all available
  \pdf{} sets). At fixed order in perturbation theory, this introduces
  an uncertainty in the bottom-quark \pdf{}s which should, however, be
  small for $\mhiggs\sim\muF\gg \mu_\text{thr}\sim \mbottom$.

  In fact, until recently, modern \pdf{} sets always assumed a
  five-flavour content of the proton. This led to an inconsistency in
  the \fs{4} calculation. With the return of four-flavour \pdf{}
  sets~\cite{Martin:2009iq,Lai:2010vv,Ball:2011mu}, this inconsistency
  could be removed.
\item[\it Numerical phase space integration:] Phase space integration in
  the \fs{4} calculation is done numerically. In the collinear region,
  this requires high precision and is computationally very demanding. In
  the \fs{5}, the analytical result for the phase space integration is
  published and implemented in a publicly available program
  \bbhnnlo~\cite{Harlander:2003ai} which performs the convolution
  with the \pdf{}s. The calculation of the cross section for a single
  set of parameters takes of the order of seconds with \bbhnnlo.
\item[\it Disagreement between the \fs{4} and \fs{5}:] If the total
  inclusive cross section is calculated at \lo{} in both the \fs{4} and
  \fs{5}, and the factorization scale is set to $\muF=\mhiggs$, one
  finds that the numerical results disagree by a factor of the order of
  five~(see, e.g., Ref.\,\cite{Kramer:2004ie}). As already mentioned in
  Section~\ref{sec::intro}, it was argued that in the \fs{5}, the
  ``proper'' choice of the factorization scale is $\muF\sim
  \mhiggs/4$. In fact, the \fs{4} and \fs{5} results agree very well in
  this case. The \nnlo{} calculation in the \fs{5} shows a rather mild
  dependence on $\muF$, and indeed confirms that $\muF=\mhiggs/4$ is a
  reasonable choice, since the radiative corrections exhibit a good
  convergence behaviour for that value.
\end{description}


\section[Santander matching]{Santander matching}
Theoretically, the \fs{4} and the \fs{5} are equivalent descriptions of
the total inclusive cross section for $(b\bar b)H$ production.
However, due to the arguments listed above, differences in the numerical
predictions are observed at finite order of perturbation theory. By
making what we believe are reasonable assumptions, in this section we
suggest a way to combine both approaches.

The \fs{4} and \fs{5} calculations provide the unique description of the
cross section in the asymptotic limits $\mhiggs/\mbottom \to 1$ and
$\mhiggs/\mbottom \to \infty$, respectively. For phenomenologically
relevant Higgs-boson masses away from these asymptotic regions both
schemes are applicable and include different types of higher-order
contributions.  The matching we suggest below interpolates between the
asymptotic limits of very light and very heavy Higgs-bosons. It is
pragmatic and far from being theoretically rigorous. However, we believe
that the matching catches the essence of the arguments given in
Section~\ref{sec::discussion} and provides a prediction with an
uncertainty band which covers the physical result.

A comparison of the \fs{4} and \fs{5} calculations reveals that both are
in numerical agreement for moderate Higgs-boson masses (see Fig.\,23 of
Ref.\,\cite{Dittmaier:2011ti}).  Once larger Higgs-boson masses are
considered, the effect of the collinear logarithms $\ln
(\mhiggs/\mbottom)$ becomes more and more important and the two
approaches begin to differ. We suggest to combine the two approaches in
such a way that they are given variable weight, depending on the value
of the Higgs-boson mass.

The difference between the two approaches is formally
logarithmic. Therefore, the dependence of their relative importance
on the Higgs-boson mass should be controlled by a logarithmic term. We determine the
coefficients such that
\begin{itemize}
\item[(a)] the \fs{5} gets 100\% weight in the limit $\mhiggs/\mbottom \to \infty$\,;
\item[(b)] the \fs{4} gets 100\% weight in the limit where the logarithms
  are ``small''. There is obviously quite some arbitrariness in this
  statement. We assume here that ``small'' means $\ln (\mhiggs/\mbottom) = 2$.
  The consequence of this particular choice is that the \fs{4} and the \fs{5}
  both get the same weight for Higgs-boson masses around 100\,GeV, consistent
  with the observed agreement between the \fs{4} and the \fs{5} in this
  region.\footnote{Note that one should use the {\it pole mass} for $\mbottom$ here
  rather than the running mass, since it is really the dynamical mass
  that rules the re-summed logarithms.}
\end{itemize}
This leads to the following formula\footnote{This formula originated
  from discussions among the authors at the {\it Higgs Days at Santander
    2009} and is therefore dubbed ``Santander matching''.}
\begin{equation}
\begin{split}
\sigma^\text{matched}= \frac{\sigma^\text{\fs{4}} +
  w\,\sigma^\text{\fs{5}}}{1+w}\,,
\end{split}
\end{equation}
with the weight $w$ defined as 
\begin{equation}
\begin{split}
w = \ln\frac{\mhiggs}{\mbottom}  - 2\,,
\label{eq::t}
\end{split}
\end{equation}
and $\sigma^\text{\fs{4}}$ and $\sigma^\text{\fs{5}}$ denote the total
inclusive cross section in the \fs{4} and the \fs{5}, respectively.
For $\mbottom=4.75$\,GeV and specific values of $\mhiggs$, this leads to
\begin{equation}
\begin{split}
\sigma^\text{matched} \big|_{\mhiggs=100\,\text{GeV}} 
&= 0.49\,\sigma^\text{\fs{4}} + 0.51\,\sigma^\text{\fs{5}}\,,\\
\sigma^\text{matched} \big|_{\mhiggs=200\,\text{GeV}} 
&= 0.36\,\sigma^\text{\fs{4}} + 0.64\,\sigma^\text{\fs{5}}\,,\\
\sigma^\text{matched} \big|_{\mhiggs=300\,\text{GeV}} 
&= 0.31\,\sigma^\text{\fs{4}} + 0.69\,\sigma^\text{\fs{5}}\,, \\
\sigma^\text{matched} \big|_{\mhiggs=400\,\text{GeV}} 
&= 0.29\,\sigma^\text{\fs{4}} + 0.71\,\sigma^\text{\fs{5}}\,, \\
\sigma^\text{matched} \big|_{\mhiggs=500\,\text{GeV}} 
&= 0.27\,\sigma^\text{\fs{4}} + 0.73\,\sigma^\text{\fs{5}}\, .
\end{split}
\end{equation}
A graphical representation of the weight factor $w$ is shown in
\fig{fig::xs4f5fhks}\,(a).

Concerning the uncertainties, we suggest to add them linearly, using the
weights $w$ defined in \eqn{eq::t}.  This ensures that the combined
error is always larger than the minimum of the two individual
errors. Neglecting correlations and assuming equality of the
uncertainties in the \fs{4} and \fs{5} calculations would imply that the
matched uncertainty is reduced by a factor of $w/(1+w)$ with respect to
the common individual uncertainties.  This seems unreasonable. In our
approach the matched uncertainty would be equal to the individual ones
in this case. On the other hand, taking the envelope of the \fs{4} and
\fs{5} error bands seems overly conservative to us.

The theoretical uncertainties in the \fs{4} and the \fs{5} calculations
are obtained through $\muF$-, $\muR$-, \pdf{}-, and $\alpha_s$ variation
as described in Ref.\,\cite{Dittmaier:2011ti}. They can be quite
asymmetric, which is why the combination should be done separately for
the upper and the lower uncertainty limits:
\begin{equation}
\begin{split}
\Delta\sigma_\pm = \frac{\Delta\sigma_\pm^\text{\fs{4}}
  + w\,\Delta\sigma_\pm^\text{\fs{5}}}{1+w}\,,
\end{split}
\label{eq::error}
\end{equation}
where $\Delta\sigma_\pm^\text{\fs{4}}$ and
$\Delta\sigma_\pm^\text{\fs{5}}$ are the upper/lower uncertainty limits
of the \fs{4} and the \fs{5}, respectively.


\section{Numerical results}

We shall discuss the numerical implications of the matching
prescriptions defined in the previous section and provide matched
predictions for the inclusive cross section $pp \to (b\bar{b})H+X$ at
the \lhc{} operating at a center-of-mass energy of 7~TeV. The individual
numerical results for the \fs{4} and \fs{5} calculations have been
obtained in the context of the
\href{https://twiki.cern.ch/twiki/bin/view/LHCPhysics/CrossSections}{\it
  LHC Higgs Cross Section Working Group} with input parameters as
described in Ref.\,\cite{Dittmaier:2011ti}. Note that the cross section
predictions presented below correspond to Standard Model bottom Yukawa
couplings and a bottom-quark mass of $\mbottom = 4.75$~GeV. \susy{}
effects can be taken into account by simply rescaling the bottom Yukawa
coupling to the proper value~\cite{Dittmaier:2006cz,Dawson:2011pe}.

\begin{figure}
  \begin{center}
    \begin{tabular}{cc} 
      \includegraphics[width=.46\textwidth]{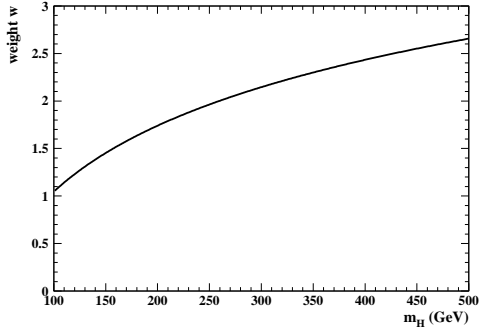} &
      \includegraphics[width=.46\textwidth]{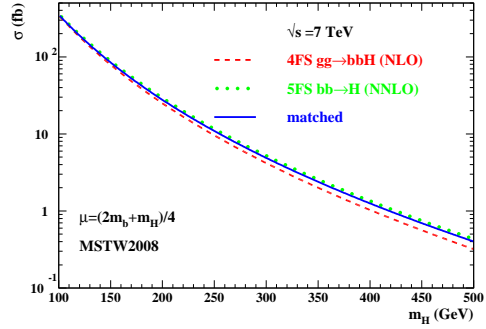} \\
      (a) & (b)
    \end{tabular}
    \parbox{.9\textwidth}{
      \caption{\label{fig::xs4f5fhks}\sloppy (a) Weight factor $w$,
        \eqn{eq::t}, as a function of the Higgs-boson mass $\mhiggs$.
        The bottom-quark pole mass has been set to $\mbottom =
        4.75$~GeV. (b) Central values for the total inclusive cross
        section in the \fs{4} (red, dashed), the \fs{5} (green, dotted),
        and for the matched cross section (blue, solid). Here and in the
        following we use the {\abbrev MSTW2008}\,\pdf{}
        set~\cite{Martin:2009iq} (\nlo{} for \fs{4}, \nnlo{} for
        \fs{5}).}}
  \end{center}
\end{figure}

\begin{figure}
  \begin{center}
    \begin{tabular}{cc}
      \includegraphics[width=.46\textwidth]{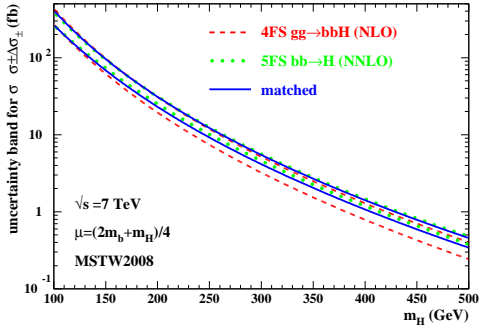} &
      \includegraphics[width=.46\textwidth]{%
        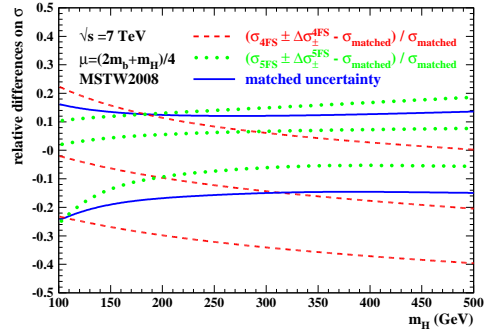}  \\
      (a) & (b)
    \end{tabular}
    \parbox{.9\textwidth}{
      \caption{\label{fig::uncertainty}\sloppy (a) Theory uncertainty
        bands for the total inclusive cross section in the \fs{4} (red,
        dashed), the \fs{5} (green, dotted), and for the matched cross
        section (blue, solid).  (b) Uncertainty bands and central
        values, relative to the central value of the matched result
        (same line coding as panel (a)).}}
  \end{center}
\end{figure}

\fig{fig::xs4f5fhks}\,(b) shows the central values for the \fs{4} and
the \fs{5} cross section, as well as the matched result, as a function
of the Higgs-boson mass. The ratio of the central \fs{4} and the \fs{5}
predictions to the matched result is displayed in
\fig{fig::uncertainty}\,(b) (central dashed and dotted line): for
$\mhiggs= 100$\,GeV, the \fs{4} and the \fs{5} contribute with
approximately the same weight to the matched cross section, with
deviations between the individual \fs{4} and the \fs{5} predictions and
the matched one of less than 5\%. With increasing Higgs-boson mass, the
\fs{4} result deviates more and more from the matched cross section due
to its decreasing weight. At $\mhiggs=500$\,GeV, it agrees to less than
20\% with the matched result, while the \fs{5} is still within 8\% of
the latter.

The corresponding theory error estimates are shown in
\fig{fig::uncertainty}. The absolute numbers are displayed in
panel\,(a), while in panel\,(b) they are shown relative to the central
value of the matched result. Up to $\mhiggs\approx 300$\,GeV, the
combined uncertainty band covers the central values of both the \fs{4}
and the \fs{5}. For larger Higgs-boson masses, the \fs{4} central value
is slightly outside this band.


%
%


%
%



\section{Conclusions}

We have reviewed the main arguments for and against following either the
\fs{4} or the \fs{5} when predicting the value of the total inclusive
cross section for bottom-quark associated Higgs-boson production. Based
on this discussion, we have suggested a way to combine the two
approaches such that their central values and uncertainties enter with
appropriate weights.



\paragraph{Acknowledgments.} 
Our thanks go to M.~Warsinsky for collaboration, discussion, and
comments, and in particular for producing the $n$-tuples which were used
for the numerics in this note; to S.~Dittmaier, M.~Spira, and W.~Kilgore
for collaboration on the original \fs{4} and \fs{5} calculations on
which the results in this note are based; to the \lhc{} Higgs Cross
Section Working Group for ideological support; and to S.~Heinemeyer
for organizing the ``Higgs Days at Santander''. All of the authors are
supported by the Helmholtz Alliance ``Physics at the Terascale''.  In
addition, MK is supported by the DFG SFB/TR9 ``Computational Particle
Physics'', RH by BMBF grant 05H09PXE, and MS by BMBF grant 05H09VFC
(Forschungsschwerpunkt FSP101 ATLAS-Experiment ``Physik auf der
TeV-Skala am Large Hadron Collider''). RH would like to thank the CERN
PH-TH group for their hospitality.




\def\app#1#2#3{{\it Act.~Phys.~Pol.~}\jref{\bf B #1}{#2}{#3}}
\def\apa#1#2#3{{\it Act.~Phys.~Austr.~}\jref{\bf#1}{#2}{#3}}
\def\annphys#1#2#3{{\it Ann.~Phys.~}\jref{\bf #1}{#2}{#3}}
\def\cmp#1#2#3{{\it Comm.~Math.~Phys.~}\jref{\bf #1}{#2}{#3}}
\def\cpc#1#2#3{{\it Comp.~Phys.~Commun.~}\jref{\bf #1}{#2}{#3}}
\def\epjc#1#2#3{{\it Eur.\ Phys.\ J.\ }\jref{\bf C #1}{#2}{#3}}
\def\fortp#1#2#3{{\it Fortschr.~Phys.~}\jref{\bf#1}{#2}{#3}}
\def\ijmpc#1#2#3{{\it Int.~J.~Mod.~Phys.~}\jref{\bf C #1}{#2}{#3}}
\def\ijmpa#1#2#3{{\it Int.~J.~Mod.~Phys.~}\jref{\bf A #1}{#2}{#3}}
\def\jcp#1#2#3{{\it J.~Comp.~Phys.~}\jref{\bf #1}{#2}{#3}}
\def\jetp#1#2#3{{\it JETP~Lett.~}\jref{\bf #1}{#2}{#3}}
\def\jphysg#1#2#3{{\small\it J.~Phys.~G~}\jref{\bf #1}{#2}{#3}}
\def\jhep#1#2#3{{\small\it JHEP~}\jref{\bf #1}{#2}{#3}}
\def\mpl#1#2#3{{\it Mod.~Phys.~Lett.~}\jref{\bf A #1}{#2}{#3}}
\def\nima#1#2#3{{\it Nucl.~Inst.~Meth.~}\jref{\bf A #1}{#2}{#3}}
\def\npb#1#2#3{{\it Nucl.~Phys.~}\jref{\bf B #1}{#2}{#3}}
\def\nca#1#2#3{{\it Nuovo~Cim.~}\jref{\bf #1A}{#2}{#3}}
\def\plb#1#2#3{{\it Phys.~Lett.~}\jref{\bf B #1}{#2}{#3}}
\def\prc#1#2#3{{\it Phys.~Reports }\jref{\bf #1}{#2}{#3}}
\def\prd#1#2#3{{\it Phys.~Rev.~}\jref{\bf D #1}{#2}{#3}}
\def\pR#1#2#3{{\it Phys.~Rev.~}\jref{\bf #1}{#2}{#3}}
\def\prl#1#2#3{{\it Phys.~Rev.~Lett.~}\jref{\bf #1}{#2}{#3}}
\def\pr#1#2#3{{\it Phys.~Reports }\jref{\bf #1}{#2}{#3}}
\def\ptp#1#2#3{{\it Prog.~Theor.~Phys.~}\jref{\bf #1}{#2}{#3}}
\def\ppnp#1#2#3{{\it Prog.~Part.~Nucl.~Phys.~}\jref{\bf #1}{#2}{#3}}
\def\rmp#1#2#3{{\it Rev.~Mod.~Phys.~}\jref{\bf #1}{#2}{#3}}
\def\sovnp#1#2#3{{\it Sov.~J.~Nucl.~Phys.~}\jref{\bf #1}{#2}{#3}}
\def\sovus#1#2#3{{\it Sov.~Phys.~Usp.~}\jref{\bf #1}{#2}{#3}}
\def\tmf#1#2#3{{\it Teor.~Mat.~Fiz.~}\jref{\bf #1}{#2}{#3}}
\def\tmp#1#2#3{{\it Theor.~Math.~Phys.~}\jref{\bf #1}{#2}{#3}}
\def\yadfiz#1#2#3{{\it Yad.~Fiz.~}\jref{\bf #1}{#2}{#3}}
\def\zpc#1#2#3{{\it Z.~Phys.~}\jref{\bf C #1}{#2}{#3}}
\def\ibid#1#2#3{{ibid.~}\jref{\bf #1}{#2}{#3}}
\def\otherjournal#1#2#3#4{{\it #1}\jref{\bf #2}{#3}{#4}}
\newcommand{\jref}[3]{{\bf #1} (#2) #3}
\newcommand{\hepph}[1]{\href{http://arxiv.org/abs/hep-ph/#1}{\tt [hep-ph/#1]}}
\newcommand{\arxiv}[2]{\href{http://arxiv.org/abs/#1}{\tt [arXiv:#1]}}
\newcommand{\bibentry}[4]{#1, {\it #2}, #3\ifthenelse{\equal{#4}{}}{}{, }#4.}


\end{document}